\documentclass[twocolumn,showkeys,preprintnumbers,floatfix,amsmath,a4paper,amssymb,nofootinbib]{revtex4}

\usepackage{amsmath}
\usepackage{graphicx}
\usepackage{dcolumn}
\usepackage{bm}
\usepackage{amssymb}
\usepackage{epstopdf}
\usepackage{xcolor}
\usepackage{comment}

\usepackage{xcolor}
\usepackage[colorlinks = true,
            linkcolor = blue,
            urlcolor  = blue,
            citecolor = blue,
            anchorcolor = blue]{hyperref}

\begin{document}
\title{A hyperbolic cell cycle law for early embryonic developmental timing}
\author{Adri\'an Aguirre-Tamaral$^{1}$, Johanna Royer$^{1,6}$, Magdalena Schindler-Johnson$^{2,3,6}$,  Jun-Ru Lee$^2$, Daniel R. Amor$^{4,5}$,  Nicoletta I. Petridou$^{2*}$ and  Bernat Corominas-Murtra$^{1}$}
\thanks{bernat.corominas-murtra@uni-graz.at\\$^*$nicoletta.petridou@embl.de}
\affiliation{
$^1$Department of Biology, University of Graz, Universit\"atsplatz 2, 8010 Graz, Austria;
$^2$European Molecular Biology Laboratory, Meyerhofstrasse 1, 69117, Heidelberg, Germany;
$^3$Collaboration for joint PhD Degree between EMBL and Heidelberg University, Faculty of Biosciences, Heidelberg, Germany;
$^4$LPENS, D\'epartement de physique, \'Ecole normale sup\'erieure, Universit\'e PSL, Sorbonne Universit\'e, Universit\'e Paris Cit\'e, CNRS, Paris, France;
$^5$IAME, Universit\'e de Paris Cit\'e, Universit\'e Sorbonne Paris Nord, INSERM, Paris, France.
$^6$Equal contribution
$^*$Author for correspondence
}
\keywords{Cell Cycle, Cell Cycle Length elongation, Hyperbolic growth, Singularity, Cleavage period, Developmental timing, Embryogenesis, Chemical Kinetics}
\begin{abstract}
Across metazoans, early embryos exhibit a strikingly conserved slowing down of their cell duplication speed, despite widely varying developmental paces and underlying molecular mechanisms. Here we show that this common behavior arises because early development unfolds along a biochemical rather than a chronological timescale, resulting from the coupling of finite maternal resource consumption to the Michaelis–Menten-like kinetics governing the rates of the biochemical reactions involved in cell duplication. This leads to a hyperbolic growth of the Cell Cycle Length (CCL), approaching a mathematical singularity, which would correspond to developmental arrest. Data from a wide range of organisms ---cnidarians, nematodes, arthropods, molluscs, echinoderms, tunicates, amphibians, and fish--- collapse on a single curve, quantitatively capturing not only a universal CCL dynamical behaviour, but also key hallmarks of early metazoan development, including cell-number temporal evolution, the dependency of CCL on cell size, and, remarkably, gastrulation timing at the predicted singularity. Crucially, experimental modulation of resource availability and consumption rates validate the model and further demonstrate that a source of heterochrony in early development is an altered biochemical timescale of resource depletion. Overall, this work reveals resource consumption rates as a fundamental mechanism driving developmental timing in early embryogenesis across species.

\end{abstract}
\maketitle

\begin{figure*}[ht!]
\begin{center}
\includegraphics[width=17.7cm]{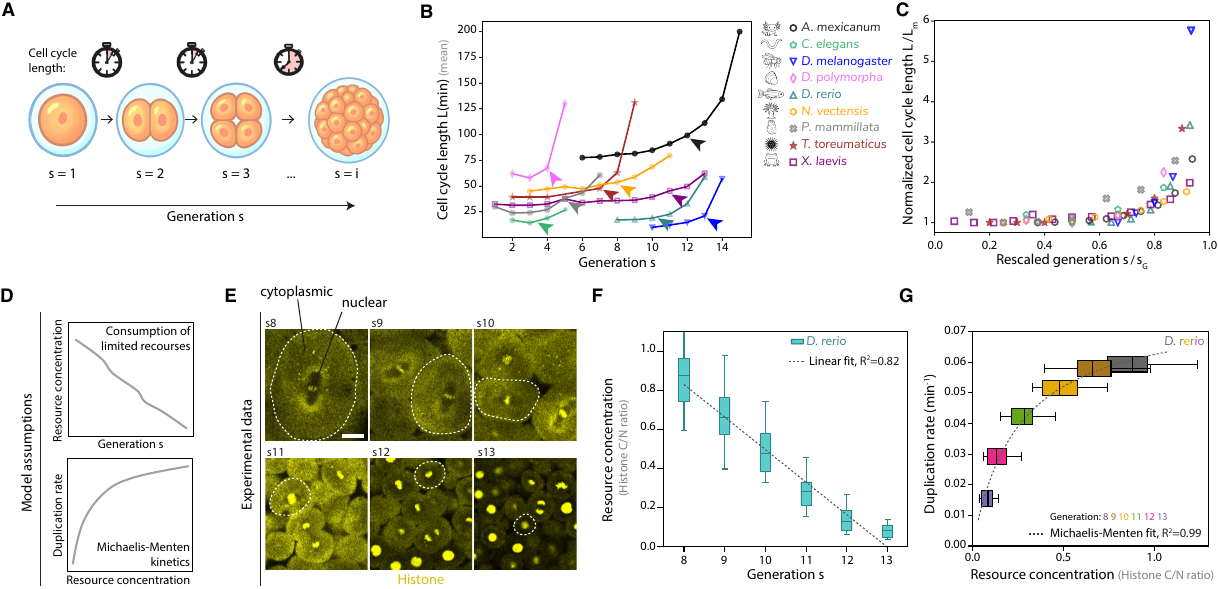}	
 \caption{
{\bf Cell cycle length (CCL) elongation during early embryonic development.} A) Schematic diagram of early embryogenesis depicting the rapid, reductive cell divisions (cleavages), followed by the Mid-Blastula transition (MBT) in which a marked elongation of the cell cycle length is observed. B) Plot of CCL across multiple species, with varying onset of MBT times (arrowheads) and overall durations. C) Normalized CCLs from the species shown in (B) to the fastest cell cycle (y-axis) and generation at gastrulation (x-axis) revealing a coinciding pattern across species. D) Model assumptions: (top) depletion of finite maternal resources, acting as substrate for the chemical reactions underlying cell duplication; and (bottom) the reaction rate depends on substrate concentration following the Michaelis-Menten equation --- see eq. (\ref{eq:MM}). E) Exemplary confocal images of H2B-transgenic embryos during cleavages showing cytoplasmic histone depletion. (Scale bar= 30 $\mu m$) 
F) Histone cytoplasmic-to-nucleus ratio intensity during the cleavage period as an example of substrate depletion, showing a linear decaying trend. (n=150 cells from N=3 embryos) G) Zebrafish cell duplication rate (inverse of CCL) as a function of the normalized histone cytoplasmic-to-nucleus ratio intensity during cleavage divisions, following a Michaelis–Menten-like dependence --- see dashed M–M fit (mean per division cycle over n=664 cells from N=10 embryos).
}
\label{Fig:Fig1}	
\end{center}		
\vspace{-8mm}
\end{figure*}
The onset of Metazoan development is characterized by a series of rapid and synchronous cell divisions, known as cleavages \cite{Newport:1982, Foe:1993, Kimmel:1995, Farrell:2014, Wolpert:2007, schindler2024collective}. These divisions are commonly reductive, meaning that daughter cells are approximately half the size of the mother cell (Fig. \ref{Fig:Fig1}A).  At a certain generation upon fertilization, cell cycles elongate, marking the phenomenon of the so-called Mid-Blastula Transition (MBT) \cite{Farrell:2014, Keller:2008, Mendieta:2013} (Fig. \ref{Fig:Fig1}B, arrowheads). The potential underlying mechanisms triggering cell cycle length (CCL) elongation range from resource limitation, such as depletion of replication factors, histones, or dNTPs \cite{collart2013titration,collart2017chk1,amodeo2015histone,chari2019histone,shindo2019dynamics,shindo2021excess,vastag2011remodeling,djabrayan2019metabolic,liu2019link} and exhaustion of energy reservoirs (e.g., metabolites or lipid droplets) \cite{dutta2017zebrafish,kilwein2023drosophila}, to the activity of cell cycle regulators \cite{heim2016regulation,zhang2016cell,brantley2021cell}, lineage-dependent effects, and zygotic genome activation \cite{farrell2012embryonic,farrell2013mechanism,di2013posttranslational}. The diversity of the molecular mechanisms triggering MBT led to the conclusion that the regulation of cell cycle elongation may be highly species-dependent.

However, CCL elongation is a deeply conserved feature of early metazoan development, extensively reported for species developing {\em ex utero} \cite{o2004embryonic}. Indeed, analysis of already existing data reveals that the CCLs initially elongate slowly, but suddenly they show an explosive, faster than exponential growth (Fig. \ref{Fig:Fig1}B, Supplementary Fig. 1), raising the question whether cell cycle elongation can display common dynamical patterns despite the species-specific regulation. Moreover, a simple normalization operation to the shortest CCL (y-axis) and to the generation of the maximum CCL (x-axis) for each species reveals a strikingly consistent pattern (Fig. \ref{Fig:Fig1}C), in which a large group of species collapse to the same temporal dynamics. These hidden regularities strongly suggest that the developmental timing of early embryos may be governed by fundamental constraints operating at a deeper level than the specific distinct molecular routes triggering CCL elongation in different species.
%

{\bf Resource depletion kinetics underlies CCL dynamics}.
The search for generic mechanisms underlying CCL elongation should ultimately be rooted in fundamental properties of biochemical reactions driving cell duplication.  
Crucially, the development of early embryos relies on the consumption of a fixed amount of maternal supplies that are enclosed in the fertilized oocyte, where there is very limited synthesis of new resources  \cite{Tadros:2009, marlow2010maternal, Langley:2014, Leesch:2023}.
We hypothesize that the rates of the essential reactions required for cell duplication follow the basic Michaelis-Menten kinetics.
Specifically, given a maximum reaction rate $u_M$ and a kinetic constant $K_m$, the Michaelis-Menten equation for the reaction rate $u$ of a given chemical reaction as a function of the underlying substrate concentration $c$ reads \cite{michaelis2013kinetics, fromm2012essentials}:
\begin{equation}
u=\frac{u_Mc}{K_m+c}\quad.
\label{eq:MM}
\end{equation}
In microbial ecology, the above equation is known within {\em consumer-resource models} \cite{armstrong1980competitive,CHESSON199026} as {\em Monod growth} \cite{Monod:1949},  providing a mechanistic link between nutrient concentration $c$ and replication rates $u$.
In consequence, we expect a gradual depletion of maternal resources (playing the role of the substrate $c$) through the successive division rounds (Fig. \ref{Fig:Fig1}D, top panel). 
According to eq. (\ref{eq:MM}), the particular way the substrate is depleted will lead to a predictable reduction of the rates of chemical reactions underlying the division rate, $u$ (Fig. \ref{Fig:Fig1}D, bottom panel), thereby elongating the CCL.

To test this hypothesis, we quantified maternal resources previously reported acting as titration factors for CCL elongation \cite{amodeo2015histone, chari2019histone, shindo2019dynamics, shindo2021excess}. Specifically, we live imaged zebrafish (\textit{Danio rerio}) early embryos (from 128 cell-stage until $\sim 4000$ cell-stage) expressing a fluorescent version of the histone protein H2B and evaluated histone consumption (Fig. \ref{Fig:Fig1}E). Early embryos contain maternally deposited histones that are not yet incorporated in the DNA, remaining in the cytoplasmic pool. However, as development unfolds, the DNA content is steadily increasing due to the sequential divisions, and cytoplasmic histones are being consumed to accompany the increasing DNA amount.
By quantifying the cytoplasmic to nucleus (C/N) ratio of histone intensity we observe an approximately linear decay in histone availability over cell generations, in agreement with previous reports \cite{chari2019histone} (Fig. \ref{Fig:Fig1}F). Importantly, its relation to cell duplication speed (inverse of CCL) matches the behavior expected from the Michaelis-Menten kinetics (Fig. \ref{Fig:Fig1}G), in which histones would act as a limiting substrate. Although other limiting factors beyond histone availability can exist, this analysis shows that, in general, maternal resources required for cell duplication are gradually consumed during early development. 
$\\$

{\bf Deriving a hyperbolic CCL elongation law}.
We then sought to derive a generic form of the cell duplication rate. To this end, we consider, as a proxy for CCL, the reaction rate associated with generating the required products for cell duplication under resource/substrate depletion across division rounds $s$. We consider any given resource with initial concentration of $c_0$, and we model the decrease of substrate concentration up to generation $s$ via a generic function $f(s)$. 
For $f$ to be as general as possible, we assume that i) $f(0)=0$ i.e., no resource has been consumed at the starting point, and that ii) $f(s)$ is a {\em smooth, monotonously increasing, unbounded positive function} of $s$. This mathematical function reflects biologically the dynamics of maternal resource consumption, which is gradual without instantaneous massive consumption events (i.e., {\em smooth}) and  constantly increasing due to the absence of, or very little, resource replacement (i.e., {\em monotonously increasing}). Finally, we consider $f(s)$ to be {\em unbounded}, because within a finite time, if no synthesis of new resources overcomes the depletion, all resources are exhausted, no matter how large the initial concentration is. With this definition we aim to grasp any consumption dynamics taking place in the early embryo when the synthesis of new components cannot overcome the depletion of resources. 

Within this scheme, the concentration of resources after the first $s$ generations, $c(s)$, is given by:
\[
c(s)=c_0-f(s)\quad.
\]
If we plug the above expression into eq. (\ref{eq:MM}), we obtain an abstract expression accounting for the dependency of the duplication rate on the generation $s$ under the depletion of resources described above:
\[
u(s)=\frac{u_M (c_0-f(s))}{K_m+ (c_0-f(s))}\quad.
\]
The CCL, further referred to in the equations as $L$, is the time spent for a generation until the next division round, which is, precisely, the inverse of the reaction rate, i.e. $L(s)=1/u(s)$. Therefore, by defining the constant $L_\mu$ as $L_\mu=1/u_M$,  the general evolution of the CCL over a generic, limiting substrate that is depleted throughout the division rounds $s$, $L(s)$, reads:
\begin{equation}
L(s)=\frac{K_m}{u_M(c_0-f(s))}+L_\mu\quad.
\label{eq:Lsfs}
\end{equation}
The above expression characterizes a hyperbolic growth with a singularity reachable in a finite number of division rounds. In particular, at:
\begin{equation}
s^*=f^{-1}(c_0)\quad,
\label{eq:t_G}
\end{equation}
all the resources are exhausted and the CCL goes to infinity, i.e., the cells do not duplicate anymore. The specific form of $f(s)$ defines the speed at which the dynamics approaches the singularity implying that, in case no other mechanisms enter into play to restore resource depletion, the system would be unable to continue the progression of the developmental program.

\begin{figure*}[ht!]
\begin{center}
\includegraphics[width=18cm]{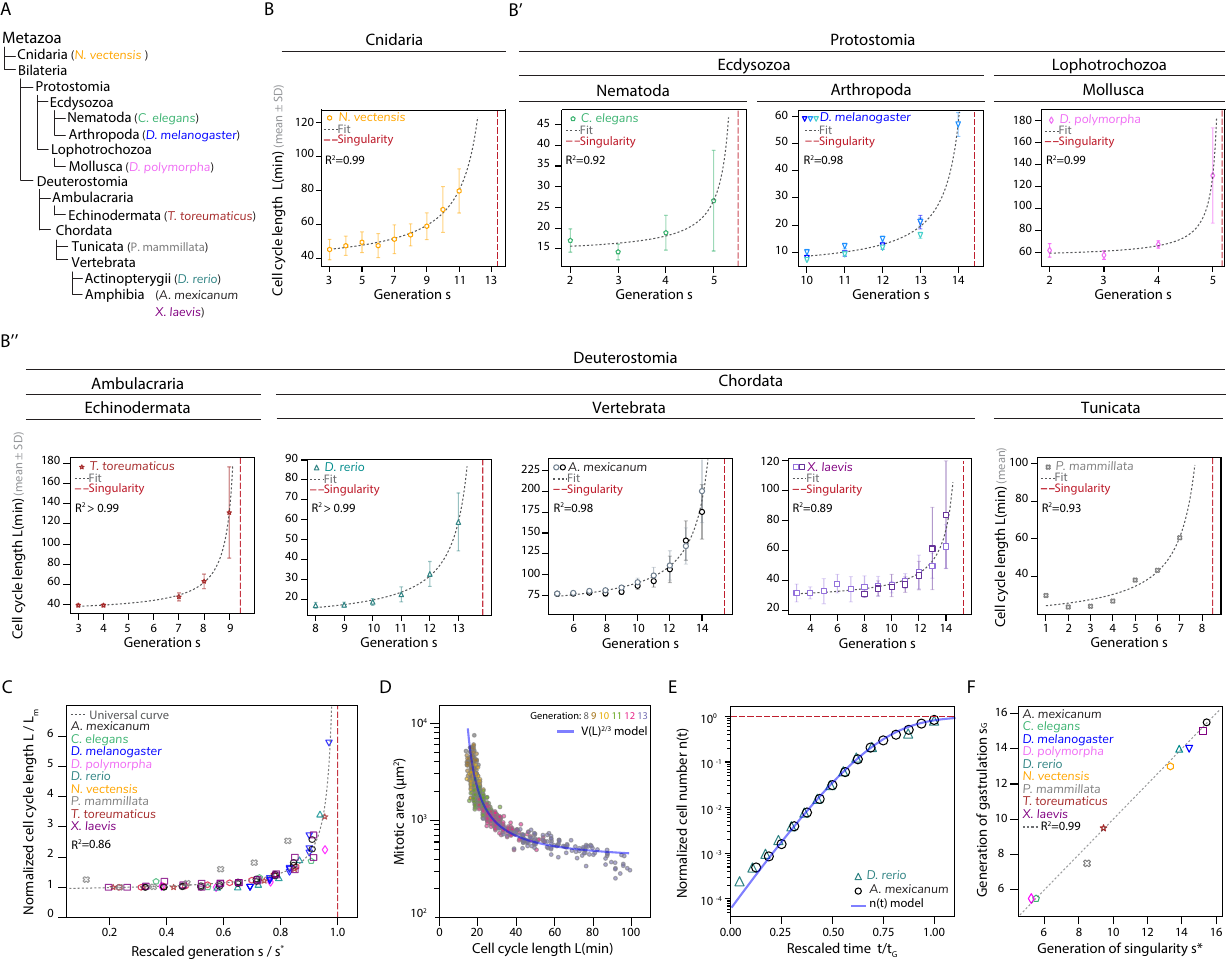}	
    \caption{{\bf Early development timing across several metazoa exhibits a hyperbolic elongation of the CCL.}
    A) Metazoan phylogenetic tree showing the phyla for which experimental data were available for quantification of CCL growth. 
    B) CCL progression during the cleavage period (cell generations) across several eumetazoan species developing {\em ex-utero}. All species display hyperbolic elongation patterns (dashed curves). Red dashed lines indicate the estimated position of the singularity $s^*$ for each species. C) Rescaling of the CCL to each species' observed minimum value $L_m$, as a function of the generation $s$ to its asymptote $s^*$, revealing a convergent behavior across eumatozoa. This universal trend is fully consistent with the hyperbolic elongation predicted by eq. (\ref{eq:rescaled}) (dashed curve). The red dashed line indicates the singularity, which is located at $1$ after rescaling.
    D) Experimental mitotic area (colored dots) as a function of CCL in wild-type zebrafish embryos, compared with the prediction of eq. (\ref{eq:Lv}) (blue line). (n=506 cells from N=8 embryos) E) Number of cells as a function of chronological time in axolotl (Exp points from \cite{hara1977cleavage}) and zebrafish WT embryos (Exp points from \cite{kimmel1995stages, jones2025quantitative}) against the prediction provided in eq.(\ref{eq:n(t)}) (blue line). F) Scatter plot of the generation corresponding to the singularity $s^*$ as predicted by mathematical fits versus the generation marking the onset of gastrulation from literature provided values. (Dashed lines correspond to data fits, while continuous lines indicate model predictions).}
\label{Fig:Fig2}	
\end{center}		
\vspace{-8mm}
\end{figure*}
$\\$

{\bf Universal behavior of the CCL elongation}.
To test whether the derived hyperbolic growth matches the evolution of the CCL observed in metazoans, we considered a linear depletion of resources per generation i.e., $f(s)\propto  s$, in agreement with the experimental data obtained from zebrafish embryos where histone availability may act as a limiting resource (Fig. \ref{Fig:Fig1}F). In consequence, the CCL elongation across generations $s$, depicted in its general form in eq. (\ref{eq:Lsfs}), acquires the following, simpler expression:
\begin{equation}
L(s)=\frac{\Lambda}{s^*-s}+L_\mu\quad,
\label{eq:Lt}
\end{equation}
with constant $\Lambda=K_m/u_M$. We tested the validity of this theoretical prediction in representatives of several major phyla in the evolutionary tree of metazoans (Fig. \ref{Fig:Fig2}A):  Chordates (zebrafish, axolotl \cite{hara1977cleavage}, frogs \cite{satoh1977metachronous, clute1995regulation}, Ascidians \cite{mcdougall2011cell}), Nematodes \cite{arata2015power, bao2008control}, Arthropodes \cite{foe1983studies, sung2013number, henry2022sponge}, Molluscs  \cite{luetjens1995multiple}, Echinoderms  \cite{masuda1984asynchronization} and Cnidarians \cite{fritzenwanker2007early}). All species exhibit an almost perfect match with the predicted hyperbolic elongation of the CCL (Fig. \ref{Fig:Fig2}B).

The scales involved in the above-derived hyperbolic growth are nevertheless dependent on the particularities of the organism under study: Some species exhibit the onset of the elongation of the CCL earlier than others, or show a more or less pronounced increase in CCL. In principle, one would expect a huge variability concerning raw size, specific maternal factors and other parameters affecting the CCL and, in particular, the amount of generations needed for exhausting these resources, i.e., the point $s^*$ at which the system would approach the singularity. However, we expect a more regular behavior across species upon rescaling, as suggested by the convergent pattern displayed in Fig \ref{Fig:Fig1}C. Seeking a universal, scale invariant behavior accounting for the CCL elongation beyond the specific species, we define the normalized CCL as a function of the rescaled generation, $\tilde{L}(s/s^*)$, as:
\begin{equation}
\tilde{L}(s/s^*)=\frac{\Lambda'}{L_m}\left(1-\frac{s}{s^*}\right)^{-1}+\frac{L_\mu}{L_m}\quad,
\label{eq:rescaled}
\end{equation}
where $L_m$ is the minimum value of the CCL observed in the particular species' experimental data under study, and $\Lambda'$ a constant, defined as $\Lambda'=\Lambda/s^*$.
Strikingly, we observe strong agreement (mean relative error $< 10\%$ and $R^2=0.86$) with the "universal" curve $\tilde{L}$ vs $s/s^*$ proposed in eq. (\ref{eq:rescaled}), valid for all considered species (Fig. \ref{Fig:Fig2}C). 
The match between CCL dynamics in early metazoan embryos to the predicted hyperbolic behavior suggests that early developmental rates follow a common, species-independent principle: Cell duplication proceeds on a biochemical timescale constrained by the Michaelis–Menten kinetics within the context of depletion of finite maternal resources.
$\\$

{\bf Predicting hallmarks of early embryogenesis}.
The early embryonic cell cycles do not only appear conserved across metazoans with respect to their dynamics, but also exhibit commonalities in additional physiological traits. For instance, cell cleavages display a conserved correlation between CCL and cell size \cite{kane1993zebrafish, arata2015power}, and they chronologically precede the onset of embryo-scale morphogenesis, gastrulation.  
Having a theoretical framework capturing the temporal dynamics of early metazoan embryogenesis prompted us to explore if one could quantitatively predict, without any additional fitting operation, hallmarks of early developmental timing.

It has long been phenomenologically established that cell volume and CCL bear an inverse correlation \cite{kane1993zebrafish, arata2015power}. Such behavior suggests that cell volume is proportional to the net resource availability. Given that the overall volume of the embryonic cell mass does not change significantly during the cleavage period, but the overall number of cells grows exponentially, one can therefore assume an exponential decrease of the cell volume. Under this assumption, and considering the evolution of the CCL as given by eq. (\ref{eq:Lt}), we arrive at the following dependency of the cell volume $v$ with respect to the CCL $L$ ---for details on the derivation, see SI:
\begin{equation}
v(L)=v_0 e^{\frac{1}{\gamma}\left[\Lambda/(L-L_\mu)-s^*\right]}\quad,
\label{eq:Lv}
\end{equation}
with $v_0$ the volume of the initial fertilized cell and $\gamma=1/\log2$, in the case of strict conservation of the volume of the overall cell mass. The above prediction displays a very good agreement against real observations for zebrafish embryonic cleaving cells (Fig. \ref{Fig:Fig2}D).

Knowing the duplication speed, we can derive the cell number evolution over chronological time. To address this question we need first to notice that, from eq. (\ref{eq:Lt}), we can approximate the time spent from fertilization to generation $s$, $t(s)$, as the sum of the duration of the $s$ previous CCLs which, under a continuous setting, reads:
\[
t(s)=\int_0^sL(s)ds\quad.
\]
By inverting $t(s)$ we obtain $s(t)$, the division round the embryo is expected to be in at (chronological) time $t$.  Since each division event implies a multiplication of the amount of cells by a factor $2$,  the number of cells is approximated by the exponential of the generation $s$. In consequence, exponentiating by $s(t)$  leads to the estimation of the amount of cells in the blastula as a function of time, $n(t)$ ---see SI for details. The specific functional form of $n(t)$ reads:
\begin{equation}
n(t)=e^{\frac{s^*}{\gamma} \left(1 - \frac{1}{\eta} W\left(\eta e^{\eta-t/\Lambda}\right) \right)}\quad,
\label{eq:n(t)}
\end{equation}
where $W(. . .)$ stands for the {\em Lambert $W$-function} \cite{abramowitz1948handbook} and $\eta = s^* L_\mu/\Lambda$ is a constant obtained from the parameters fitted in the CCL exploration reported in Fig \ref{Fig:Fig2}B, B', B'', and finally, $\gamma=1/\log 2$, as above. Interestingly, as \( t \to \infty \), $W\to W(0)=0$, and the number of cells approaches a horizontal asymptote, representing the carrying capacity of the system under the initial available resources, i.e., $n(\infty)=e^{s^*/\gamma}=2^{s^*}$, if no further synthesis was performed. We confronted the prediction provided by eq. (\ref{eq:n(t)}) with available data from zebrafish \cite{Kimmel:1995,jones2025quantitative} and axolotl \cite{hara1977cleavage} showing a very good agreement (Fig. \ref{Fig:Fig2}E).  In summary, based on the assumptions of the model, we derive the specific mathematical form relating chronological time and cell numbers along the early embryonic period.

Given that our theoretical model captures key features of early developmental progression, we further explored if it can predict the first embryonic morphogenetic event, gastrulation. Specifically, since cell cycles slow down by the onset of gastrulation, we asked whether the singular point $s^*$ predicted by the theory could actually be predictive of the point at which gastrulation starts. We found a striking correspondence between the inferred $s^*$ and the reported gastrulation generation across the studied eumetazoan branches (Fig. \ref{Fig:Fig2}F) suggesting that gastrulation occurs at the predicted singularity i.e., the point at which resources are exhausted and development would arrest \cite{Kimmel:1995, kane1992mitotic, Farrell:2014, heasman2006patterning, kitazawa2014developmentSeaUrchinGastrulation, hara1977cleavage, bucher1994gastrulation, fritzenwanker2007early}. Accordingly,  the number of cells in the embryo at gastrulation is predicted by $n(\infty)=2^{s^*}$, as defined in eq. (\ref{eq:n(t)}), identified as the limit capacity in Fig. \ref{Fig:Fig2}E. Overall, the above comparison of theoretical predictions against experimental data shows that hallmarks of early metazoan embryogenesis, including cell cycle elongation, cell size-CCL dependency and gastrulation timing are highly conserved due to fundamental biochemical constraints in maternal resource depletion. 
$\\$

\begin{figure*}
\begin{center}
\includegraphics[width=18cm]{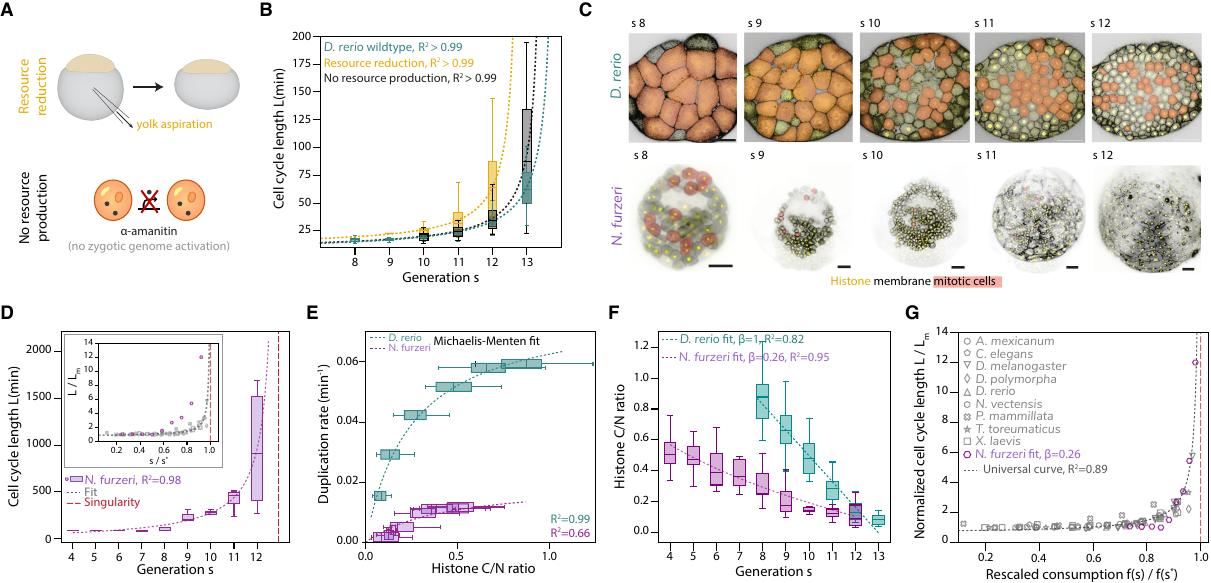}	
    \caption{
    {\bf Resource concentration and consumption rates capture developmental timing even in heterochronic species}
    A) Schematic illustrations of the experimental set-up used to reduce resource concentration via aspiration of maternal material (top) and to inhibit the activation of the zygotic genome and thus production of new resources (bottom). B) Plot of CCL over generations in WT, $\alpha$-amanitin-treated (inhibition of transcription and synthesis of new resources - grey data points) and resource-depleted embryos (aspiration - yellow data points) C) Representative experimental images showing the developmental progression of \textit{D. rerio} (top) and \textit{N. furzeri} (bottom) across different stages, from generation $s=8$ to generation $s=12$. (Scale bars = 100$\mu m$). D) Killifish CCL elongation follows a hyperbolic fit up to the generation of diapause, where development can pause at the singularity prior to gastrulation; however, this fit does not collapse onto the rescaled universal curve --- see inset, grey data points indicate the species studies in Fig. \ref{Fig:Fig2}. (Exp. data n=14 cells from 5 embryos).  E) Duplication rate in killifish (purple data) follows Michaelis-Menten kinetics as in the case of zebrafish (turquoise data). (Data from \textit{D. rerio}: n=150 cells from N=3 embryos, \textit{N. furzeri}: n=10 cells from 4 embryos.)  F) Histone cytoplasm-to-nucleus ratio in killifish follows a sublinear, power-law, decay (purple data) $f(s)\propto s^{\beta}$, with $\beta\approx 0.26$, differing from the linear trend observed in other species, such as zebrafish (turquoise data). (Data from \textit{D. rerio}: n=664 cells from N=10 embryos, \textit{N. furzeri}: n=10 cells from 4 embryos.) G) Assuming the resource consumption dynamics following $f(s)\propto s^\beta$ experimentally inferred, the killifish data (purple) collapse onto the generalized universal curve described by eq. (\ref{eq:rescaledbeta}). }
\label{Fig:Fig3}	
\end{center}		
\vspace{-7mm}
\end{figure*}

{\bf Early developmental timing is determined by resource availability and chemical kinetics}.
We further aimed to validate the mathematical approach provided above by challenging its assumptions via experimentally interfering with resource availability.
According to the proposed theoretical framework, the initial concentration of resources $c_0$ defines the location of the mathematical singularity $s^*$, and that full exhaustion of resources drives the system into the singularity. 
To test the above hypotheses, we first experimentally addressed whether an additional depletion of resources right after fertilization would result in an earlier elongation of the CCL. Still, the hyperbolic growth pattern should be conserved, albeit with different parametrization. To that end, we performed aspirations in the zygote of zebrafish one-cell stage embryos, right after fertilization (Fig. \ref{Fig:Fig3}A). The fertilized oocyte, composed of a mixed yolk and cytoplasm acts as a reservoir providing maternal resources. Therefore, if we remove part of the oocyte mass, we can reduce the available maternal resources. We analyzed the evolution of the CCL of embryos under aspiration and, as expected the CCL elongation started, on average, around 1.5 generations earlier than in the wildtype, still keeping the hyperbolic profile (Fig. \ref{Fig:Fig3}B). The acceleration of the CCL elongation observed is in agreement with similar experimental manipulations of resource availability performed in \textit{Xenopus} embryos \cite{clute1995regulation}, further supporting and validating the change in the singular point predicted by the model (Supplementary Fig. 2A). 

Without resource restoration or synthesis by the time of gastrulation, early embryos would reach the singularity and thus developmentally arrest. However, this is not the case since, in wildtype organisms, at some point during the early cleavages, the zygotic genome is activated and resources start being synthesized \cite{langley2014new}. 
To test if this is indeed the mechanism preventing embryos from reaching the singularity, we blocked zygotic genome activation in zebrafish via injection of $\alpha$-amanitin, an RNA polymerase inhibitor, thereby blocking the synthesis of new resources (Fig. \ref{Fig:Fig3}A). We observed an aberrant elongation of the CCL, occurring only in the generation that matches the mathematical singularity (Fig. \ref{Fig:Fig3}B), despite synthesis being already inhibited from MBT onward (Supplementary Fig. 2B). This shows that indeed the singularity $s^*$ corresponds to the full maternal resource exhaustion and further suggests that the default of early embryo development is to effectively {\em fall} into the singularity. Therefore, the activation of the zygotic machinery, encoded in the developmental program, allows the embryo to escape from the bottleneck defined by the exhaustion of maternal material 
(Fig. \ref{Fig:Fig3}B).
We finally observe that if cell duplication rates are ultimately governed by enzymatic reactions, higher temperatures should increase reaction rates and, thus, lead to shorter cell-cycle lengths (CCL), and vice versa. We confirm this expected behavior by performing temperature-dependent experiments. Crucially, we also validate the theoretical prediction that the singularity $s^*$ depends primarily on the initial resource concentration $c_0$, since, even if other parameters of the hyperbola change with temperature, the location of the singularity remains at the same generation (Supplementary Fig. 2C).
$\\$

{\bf Differential resource consumption rates underlie heterochrony in early embryogenesis}.
The above results indicate that the early developmental timing may unfold at a biochemical timescale defined by the rate of resource consumption represented by the function $f(s)$. So far, we assumed the consumption dynamics to be linear, but other patterns are, by no means, excluded. To directly test this hypothesis, we evaluated the resource consumption rates and CCL elongation in species displaying heterochronies in their embryonic development, and asked if such heterochronies could be the outcome of differential resource consumption rates. To this end, we tracked the evolution of the CCL along the cleavage period of killifish {\textit{(Nothobranchius furzeri)}} (Fig. \ref{Fig:Fig3}C,D), a species that, during embryogenesis, can enter diapause, a form of heterochrony in which development arrests, also called dormancy or suspended animation \cite{fenelon2014embryonic, Renfree:2017}. We specifically focused on facultative diapause I, which occurs before gastrulation, at the dispersal phase (Fig. \ref{Fig:Fig3}C, bottom panel, generations 11,12)\cite{Wourms:1972}. Although we did not change the environmental conditions to induce full diapause I, we observe that the CCL elongation in killifish is much more pronounced, values climb much further up than all the 
other species, the cleavage stage takes proportionally longer and, crucially, the behaviour upon rescaling is qualitatively different, indicating a heterochrony (Fig. \ref{Fig:Fig3}D inset).

However, despite showing a hyperbolic growth, and a satisfactory individual fit with eq. (\ref{eq:Lt}) (Fig. \ref{Fig:Fig3}D), we observed that the behavior of the killifish CCL evolution does not match the "universal" curve described in eq. (\ref{eq:rescaled}), as the other studied metazoans do (Inset Fig. \ref{Fig:Fig3}D). 
We thus analyzed the histone consumption versus duplication rate in killifish and we found a relation compatible with the Michaelis–Menten kinetics in both cases, albeit with a slower rate in killifish (Fig. \ref{Fig:Fig3}E). To understand the reason behind the slower duplication rate, we quantified histone depletion in killifish. Although in zebrafish the C/N histone ratio is decaying linearly, in killifish we observe a slower, sublinear decay of $f(s)$, being the trend well fitted at:
\[ 
f(s)\propto s^{\beta}\quad,
\]
with $\beta\approx 0.26$ 
---see SI for details  (Fig. \ref{Fig:Fig3}F). Although the molecular mechanisms behind the differential resource consumption kinetics in killifish are yet to be found, a more conservative depletion rate is expected in systems anticipating dormancy, which were shown to exhibit different metabolic programs \cite{Bulut-Karslioglu:2016, Hu:2020}. Generalizing the framework to generic resource consumption dynamics naturally leads to a generalization of the normalized function defined in eq. (\ref{eq:rescaled}), now rescaling the variable with the general expression of the function $\tilde{L}(s/s^*)\to \tilde{L}(f(s)/f(s^*))$, as:
\begin{equation}
\tilde{L}\left(\frac{f(s)}{f(s^*)}\right)=\frac{\Lambda'}{L_m}\left(1-\frac{f(s)}{f(s^*)}\right)^{-1}+\frac{L_\mu}{L_m}\quad.
\label{eq:rescaledbeta}
\end{equation}
We observe that the CCLs of all other studied eumetazoans collapsed in the same scaling behavior because, for all of them, it turns out that a linear decay ---i.e., $f(s)\propto s$--- is a good approximation for the pace at which resources are consumed. Interestingly, we observe that the killifish CCL elongation collapses to the "universal" curve, now generalized by eq. (\ref{eq:rescaledbeta}), with the other species when the differentiated consumption of resource is taken into account (Fig. \ref{Fig:Fig3}G). Altogether, the above results demonstrate that heterochronies in early developmental timing can be captured by our mathematical model, and can be mechanistically explained solely by the consumption kinetics of finite resources.
$\\$

{\bf Discussion}.
Embryogenesis is typically compared in chronological time, yet our findings reveal a deeper regularity: Across metazoans developing {\em ex utero}, early cell cycles lengthen following a conserved hyperbolic law. We show that this behavior follows directly from the assumption that the dependency of cell duplication rates on the depletion of maternally supplied finite resources follows a Michaelis–Menten-like dynamics, rather than stage-specific regulatory programs or species-specific developmental clocks. This finding redefines the so-called MBT from a distinct developmental program that separates the early and fast cycles (pre-MBT) from the later and slow cycles (post-MBT) to a continuum regulation, where fast cycles correspond to the resource saturation regime and slow cycles to the resource limitation regime. Moreover, within this framework, diverse features of early development —including cell-number evolution, the dependency of cycle length on cell size, and the temporal positioning of gastrulation— emerge as coupled consequences of resource-limited progression. 

The above findings are aligned with recent studies showing that interspecies differences in developmental tempo and progression arise from intrinsic biochemical kinetics, including protein turnover and transcriptional delays, indicating that chronological time is secondary to intracellular reaction rates \cite{Matsuda:2020, Rayon:2020}. In these regards, we found a strikingly consistent pattern for early developmental timing across metazoans, suggesting that embryogenesis is not only paced by particular biochemical kinetics, but organized along a biochemical dynamics defined by raw resource consumption. Species-specific reaction rates determine how rapidly embryos traverse this trajectory, whereas the trajectory itself reflects a universal constraint imposed by the interplay of resource availability and consumption. This perspective also reframes heterochrony and dormancy \cite{Garcia-Ojalvo:2023}. Variations in developmental pace appear as quantitative modulations of consumption rates along the same biochemical path rather than distinct developmental programs. 

Last, the hyperbolic pattern emerging from the resource-limited chemical kinetics, predicts the existence of a dynamical singularity which quantitatively anticipates gastrulation timing. Gastrulation, as one of the most crucial developmental milestones, marks the first multicellular embryonic function, the coordination of cell fate specification and tissue motion \cite{Solnica-Krezel:2012}. The finding that gastrulation coincides with the point at which resources would be exhausted, cell cycles would diverge, and effectively development would arrest, implies that embryos need to overcome intrinsic metabolic boundaries to progress \cite{Iyer:2024, kahlon2024mitochondrial}. This indicates that resource-sensing and response pathways may underlie clonally-derived multicellular function, a strategy often found in unicellular organisms undergoing starvation-dependent aggregation and facultative multicellularity \cite{Bonner:1947, Koschwanez:2013, Kelly:2021, Schwartzman:2022, Barrere:2023}. Indeed, our analytical predictions can capture resource-limited growth dynamics for bacteria \cite{Monod:1949} and protists \cite{ajala2020assessment} (Supplementary Fig. 3). First applied to microbial growth, Michaelis–Menten kinetics ---and consequences thereof, when coupled to resource depletion--- prove equally effective in describing duplication rates in early embryos, indicating that common resource‑allocation principles may underlie both microbial physiology and early developmental progression. This suggests that evolutionary diversification of developmental strategies may arise through tuning of consumption rates that set how biochemical time is traversed. We propose the above law as a simple physical basis for both the conservation and the diversity of early developmental timing across species.
%

{\bf Acknowledgments}.
We thank Takashi Hiiragi, Dimitri F\'abreges, Guillaume Salbreaux and members of the Petridou and Corominas-Murtra group for technical advice, critical discussions and feedback on the manuscript;  We thank the Animal Facility at the Leibniz Institute on Aging for providing initial killifish embryos and the EMBL fish facility for husbandry support. This work was supported by the Weave project "Tissue material phase transitions and their role in embryo pattern formation" from the \"Osterreichischer Wissenschaftsfonds (FWF, Austrian Science Fund, I6533) to B.C-M and J. R. N.I.P. and M.S-J were supported by the European Union (European Research Council Starting Grant 101162743). B.C-M. and A. A-T. acknowledge the support of the field of excellence "Complexity of life in basic research and innovation" of the University of Graz. 

\bibliographystyle{unsrt}
\bibliography{CCL.bib}

\end{document}